\documentclass[showpacs,twocolumn,nofootinbib]{revtex4}
\usepackage{graphicx}
\usepackage{amsmath}
\usepackage{amssymb}

\begin{document}

\title{Tachyon cosmology, supernovae data and the Big Brake singularity}
\author{Z. Keresztes}
\affiliation{Department of Theoretical Physics, University of Szeged, Tisza Lajos krt
84-86, Szeged 6720, Hungary \\
Department of Experimental Physics, University of Szeged, D\'{o}m T\'{e}r 9,
Szeged 6720, Hungary}
\author{L. \'{A}. Gergely}
\affiliation{Department of Theoretical Physics, University of Szeged, Tisza Lajos krt
84-86, Szeged 6720, Hungary\\
Department of Experimental Physics, University of Szeged, D\'om T\'er 9,
Szeged 6720, Hungary\\
Department of Applied Science, London South Bank University, 103 Borough
Road, London SE1 OAA, UK}
\author{V. Gorini}
\affiliation{Dipartimento di Scienze Fisiche e Mathematiche, Universit\`a dell'Insubria,
Via Valleggio 11, 22100 Como, Italy \\
INFN, sez. di Milano, Via Celoria 16, 20133 Milano, Italy}
\author{ A. Yu. Kamenshchik}
\affiliation{Dipartimento di Fisica and INFN, via Irnerio 46, 40126 Bologna, Italy \\
L.D. Landau Institute for Theoretical Physics, Russian Academy of Sciences,
Kosygin street 2, 119334 Moscow, Russia}
\author{U. Moschella}
\affiliation{Dipartimento di Scienze Fisiche e Mathematiche, Universit\`a dell'Insubria,
Via Valleggio 11, 22100 Como, Italy \\
INFN, sez. di Milano, Via Celoria 16, 20133 Milano, Italy}

\begin{abstract}
We compare the existing observational data on type Ia Supernovae with the
evolutions of the universe predicted by a one-parameter family of tachyon
models which we have introduced recently in paper \cite{we-tach}. Among the
set of the trajectories of the model which are compatible with the data
there is a consistent subset for which the universe ends up in a new type of
soft cosmological singularity dubbed Big Brake. This opens up yet another
scenario for the future history of the universe besides the one predicted by
the standard $\Lambda$CDM model.
\end{abstract}

\pacs{98.80.Cq, 98.80.Jk, 98.80.Es, 95.36.+x}
\maketitle

\section{Introduction}

The discovery of cosmic acceleration \cite{cosm} has stimulated the study of
different models of dark energy \cite{dark} which may be responsible for
such a phenomenon. Models of dark energy include those based on different
perfect fluids, having negative pressure, on minimally and non-minimally
coupled scalar fields and on fields having non-standard kinetic terms \cite%
{kinetic,tachyons}. The latter ones include as a subclass the models based
on different forms of the Born-Infeld-type action, which is often associated
with the tachyons arising in the context of string theory \cite{string}.
Tachyonic models with relatively simple potentials were confronted with
observational data in \cite{tach-obs}. Compared to the standard Klein-Gordon
scalar field cosmological models the dynamics of tachyon models can be much
richer due to the non-linearity of the dependence of the tachyon Lagrangians
on the kinetic term of the tachyon field.

In a recent paper \cite{we-tach} a particular one-parameter family of
tachyon models was considered, which has revealed some unexpected features.
At some values of the parameter of the model a long period of accelerated
quasi-de Sitter expansion is followed by a period of cosmic deceleration
culminating, after a finite time, in an encounter with a cosmological
singularity of a new type, which was named \textit{Big Brake}. This
singularity is characterized by an infinite negative value of the second
time derivative of the cosmological radius of the universe, while its first
time derivative and the Hubble variable vanish, and the radius itself
acquires a finite value. This singularity belongs to the class of soft
(sudden) cosmological singularities \cite{soft,soft1,soft2} which have been
rather intensively studied during the last years. Here it is worth
mentioning that in the context of the scrutiny of candidates for the role of
dark energy, some other singularities attract the attention of cosmologists.
Among them a special place occupies the Big Rip singularity \cite{Rip},
arising in some models where phantom dark energy \cite{phantom} is present.
The possibility of existence of a phase of contraction of the universe,
ending up in the standard Big Crunch cosmological singularity was also
considered in the literature \cite{Crunch}. Recently, $w$-singularities were
also proposed \cite{w}.

An attractive peculiarity of the tachyon model studied in paper \cite%
{we-tach} is the fact that there the Big Brake singularity is not put in
\textquotedblleft by hands\textquotedblright , but arises naturally as a
result of the cosmological evolution, provided some initial conditions are
chosen. Therefore it is a consequence of the dynamics, rather than a pure
kinematical possibility. Such evolution leading to the Big Brake coexists
with another type of evolution describing an infinite expansion of the
universe. In other words, a small change of initial conditions can have
drastic consequences for the future of the universe. Actually, in spite of
it being somewhat exotic, we show that the cosmological model \cite{we-tach}
does not contradict observations. To this aim we compare the cosmological
evolutions predicted in \cite{we-tach} with the data coming from the
supernovae type Ia observations. We select the compatible initial conditions
by studying the backward evolution in comparison with the luminosity -
redshift diagrams for the supernovae type Ia standard(izable) candles. Then,
choosing initial conditions which are compatible at the 1$\sigma $ level
with the data, we study the forward evolution and show that a deceleration
period following the present accelerated expansion is possible, and when it
is so, we estimate how long it is expected to last.

The structure of the paper is the following. In Sec. II we introduce the
model and its basic equations; in Sec. III we find a subset of initial
conditions which are compatible with the observational data by integrating
numerically the dynamical equations backwards in time; in Sec. IV we study
numerically the cosmological evolutions for the selected initial conditions
by numerical integration forward in time. We end with some concluding
remarks.

\section{Tachyon cosmological model}

\begin{figure}[t]
\includegraphics[height=6cm, angle=360]{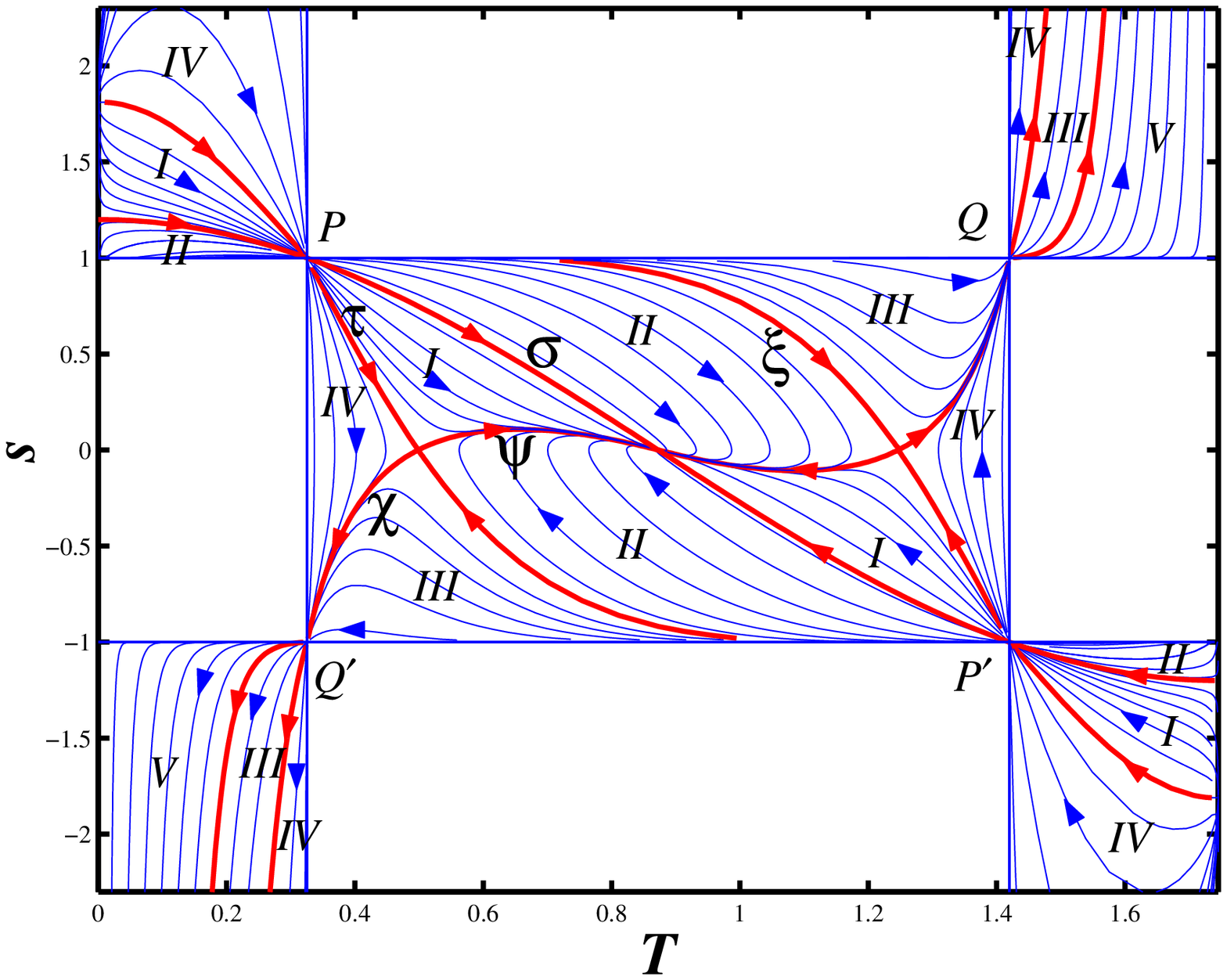}
\caption{(Color online) Phase portrait evolution for $k>0$ ($k=0.44$).}
\label{Fig1}
\end{figure}

We consider the flat Friedmann universe with the metric $%
ds^{2}=dt^{2}-a^{2}(t)dl^{2}$, filled with a spatially homogeneous tachyon
field $T$ evolving according to the Lagrangian 
\begin{equation}
L=-V(T)\sqrt{1-g_{00}\dot{T}^{2}}.  \label{Lagrange}
\end{equation}%
The energy density and the pressure of this field are, respectively 
\begin{equation}
\varepsilon =\frac{V(T)}{\sqrt{1-\dot{T}^{2}}}  \label{energy}
\end{equation}%
and 
\begin{equation}
p=-V(T)\sqrt{1-\dot{T}^{2}}.  \label{pressure}
\end{equation}%
The equation of motion for the tachyon is 
\begin{equation}
\frac{\ddot{T}}{1-\dot{T}^{2}}+3\frac{\dot{a}\dot{T}}{a}+\frac{V_{,T}}{V}=0.
\label{eq-of-motion}
\end{equation}%
We consider the following tachyon potential $V(T)$ \cite{we-tach}: 
\begin{eqnarray}
&&V(T)=\frac{\Lambda }{\sin ^{2}\left( \frac{3}{2}\sqrt{\Lambda (1+k)}%
T\right) }  \notag \\
&&\times \sqrt{1-(1+k)\cos ^{2}\left( \frac{3}{2}\sqrt{\Lambda (1+k)}%
T\right) },  \label{poten}
\end{eqnarray}%
where $\Lambda $ is a positive constant and $-1<k<1$.

Taking into account the Friedmann equation $H^{2}=\varepsilon $, where the
Hubble variable $H$ is defined as $H\equiv \dot{a}/a$, and the Newtonian
constant is normalized as $8\pi G/3=1$, we obtain the following dynamical
system: 
\begin{equation}
\dot{T}=s,  \label{system1}
\end{equation}%
\begin{equation}
\dot{s}=-3\sqrt{V}(1-s^{2})^{3/4}s-(1-s^{2})\frac{V_{,T}}{V}.
\label{system2}
\end{equation}%
When the parameter $k$ is negative, the evolution of the system (\ref%
{system1})-(\ref{system2}) is confined inside the rectangle 
\begin{equation}
-1\leq s\leq 1,  \label{rectangle}
\end{equation}%
\begin{equation}
0\leq T\leq \frac{2\pi }{3\sqrt{\Lambda (1+k)}}.  \label{rectangle1}
\end{equation}%
The system has only one critical point: 
\begin{equation}
T_{0}=\frac{\pi }{3\sqrt{\Lambda (1+k)}},\ \ s_{0}=0,  \label{crit}
\end{equation}%
which is an attractive node corresponding to a de Sitter expansion with
Hubble parameter 
\begin{equation}
H_{0}=\sqrt{\Lambda }.
\end{equation}%
All cosmological histories begin at the Big Bang type cosmological
singularity located on the upper ($s=1$) or lower ($s=-1$) side of the
rectangle (\ref{rectangle})-(\ref{rectangle1}), the individual history being
parametrized by the initial value of $T$ satisfying the inequality (\ref%
{rectangle1}). They all end up in the node (\ref{crit}).

In the case $k > 0$ the situation is more complicated. First of all, the
real potential $V$ is well-defined only in the interval 
\begin{equation}
T_3 \leq T \leq T_4,  \label{interval}
\end{equation}
where 
\begin{equation}
T_3 = \frac{2}{3\sqrt{(1+k)\Lambda}} \mathrm{arccos} \frac{1}{\sqrt{1+k}},
\label{T3}
\end{equation}
\begin{equation}
T_4 = \frac{2}{3\sqrt{(1+k)\Lambda}}\left(\pi - \mathrm{arccos} \frac{1}{%
\sqrt{1+k}}\right).  \label{T4}
\end{equation}

\begin{widetext}

\begin{figure}[ht]
\includegraphics[height=8cm, angle=270]{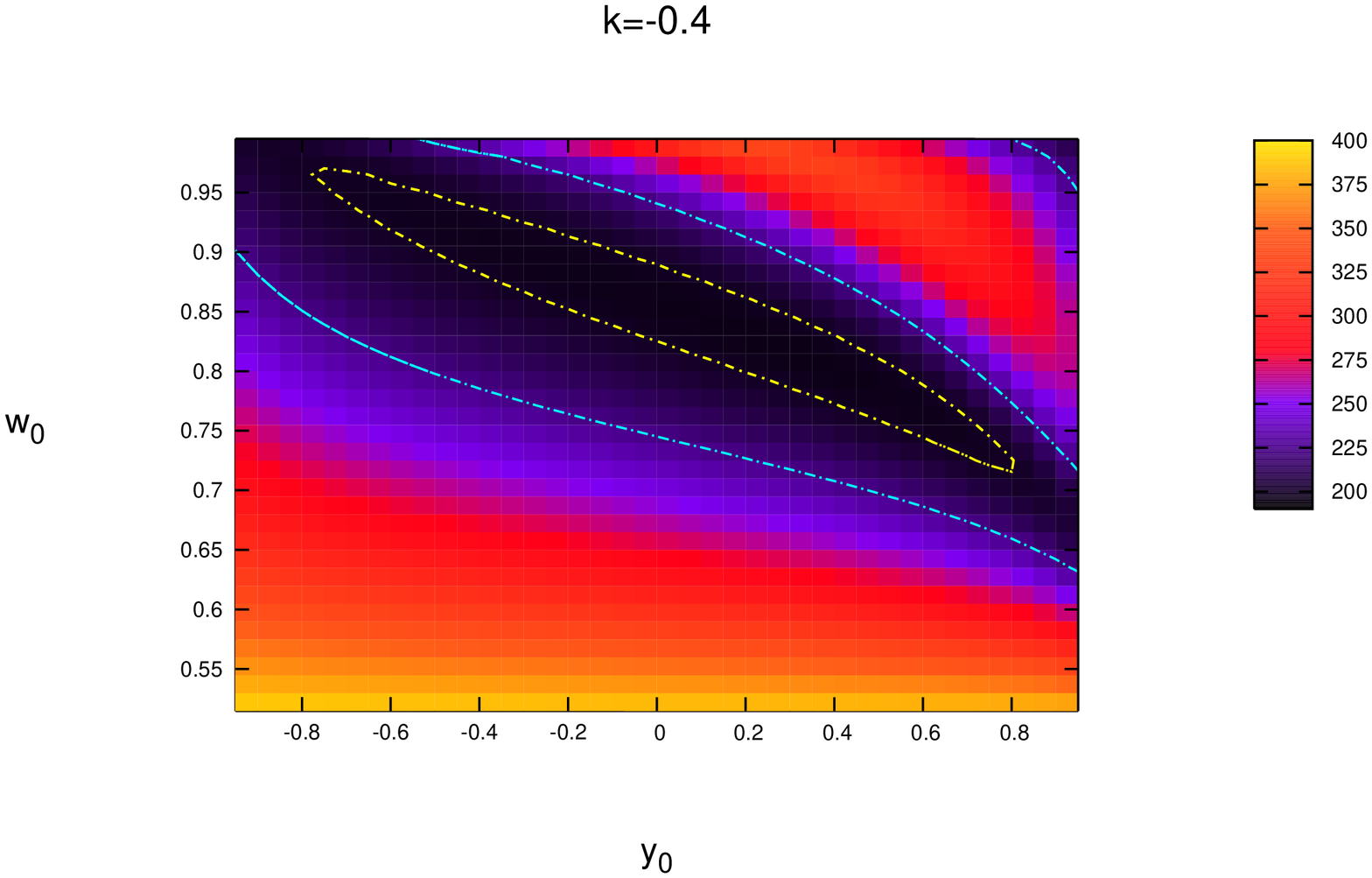} %
\includegraphics[height=8cm, angle=270]{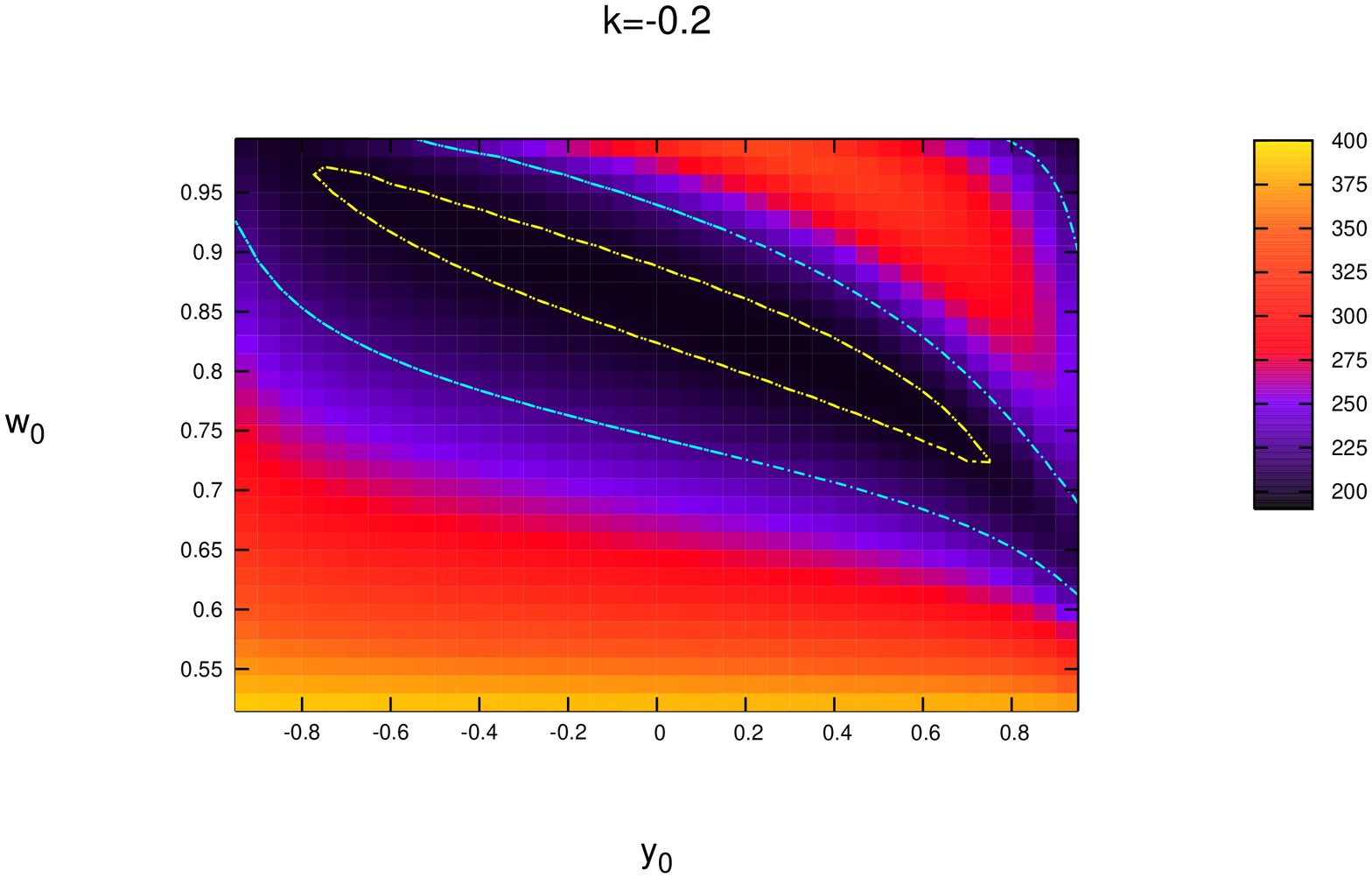} %
\includegraphics[height=8cm, angle=270]{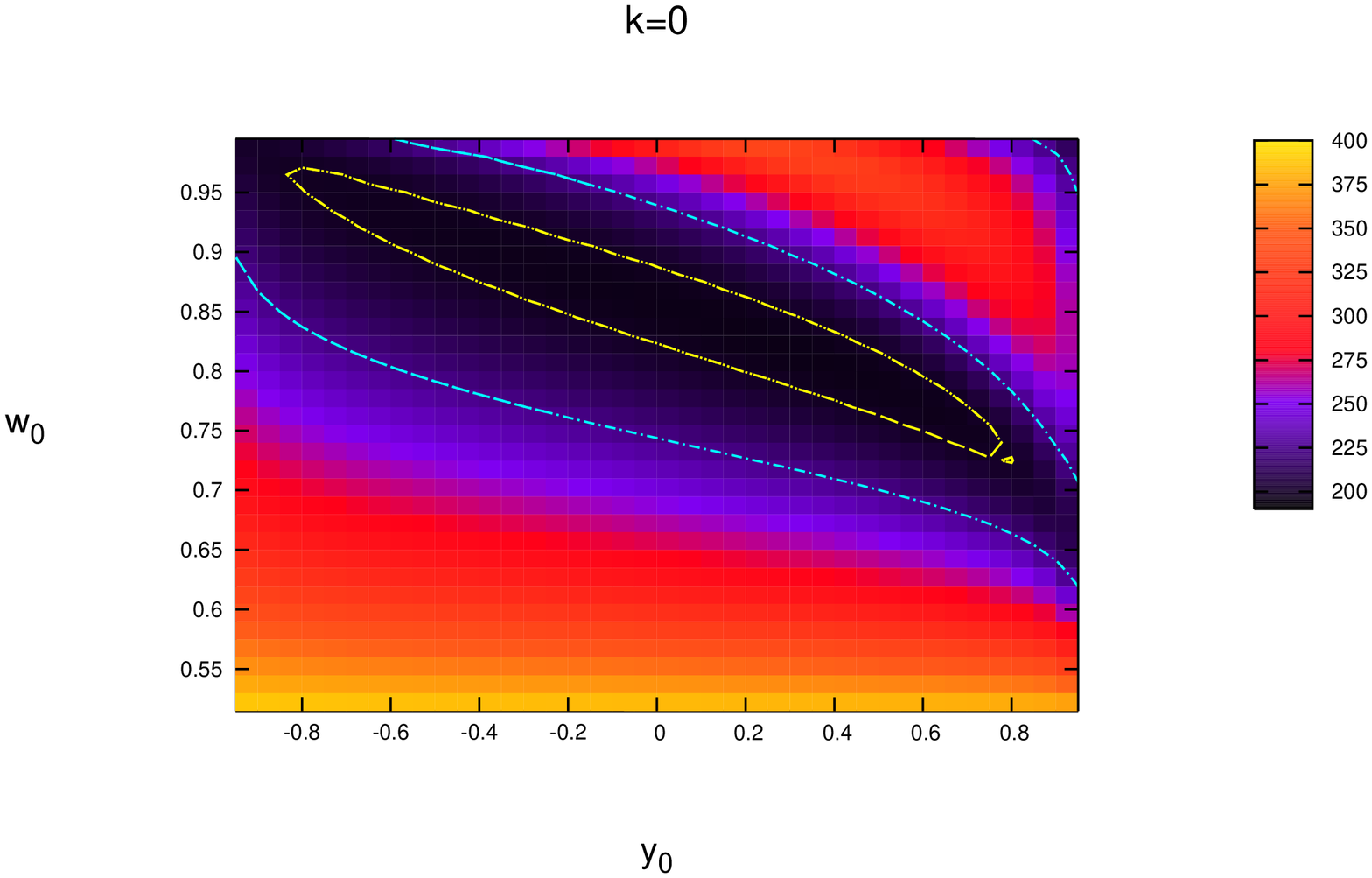} %
\includegraphics[height=8cm, angle=270]{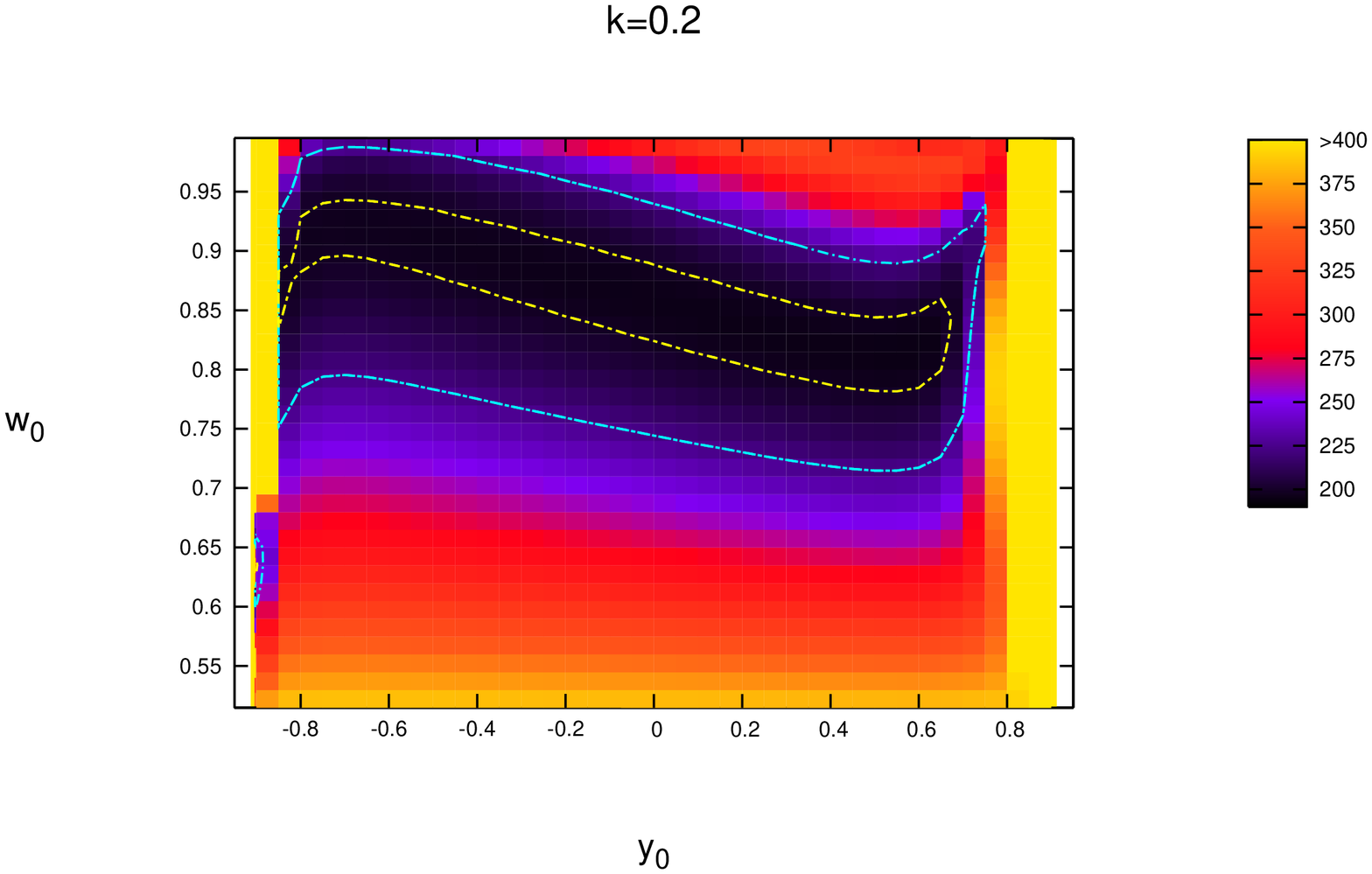} %
\includegraphics[height=8cm, angle=270]{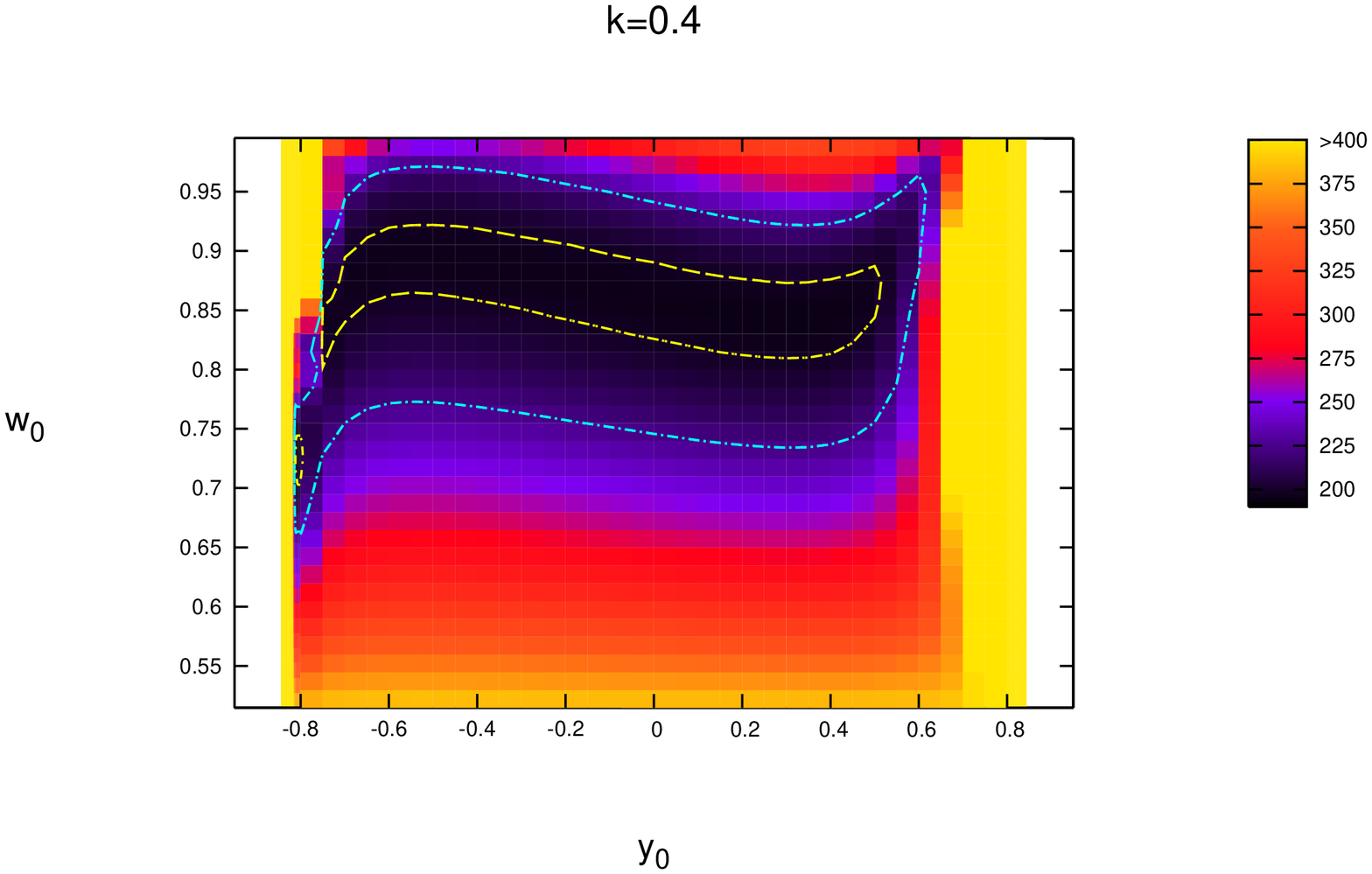} %
\includegraphics[height=8cm, angle=270]{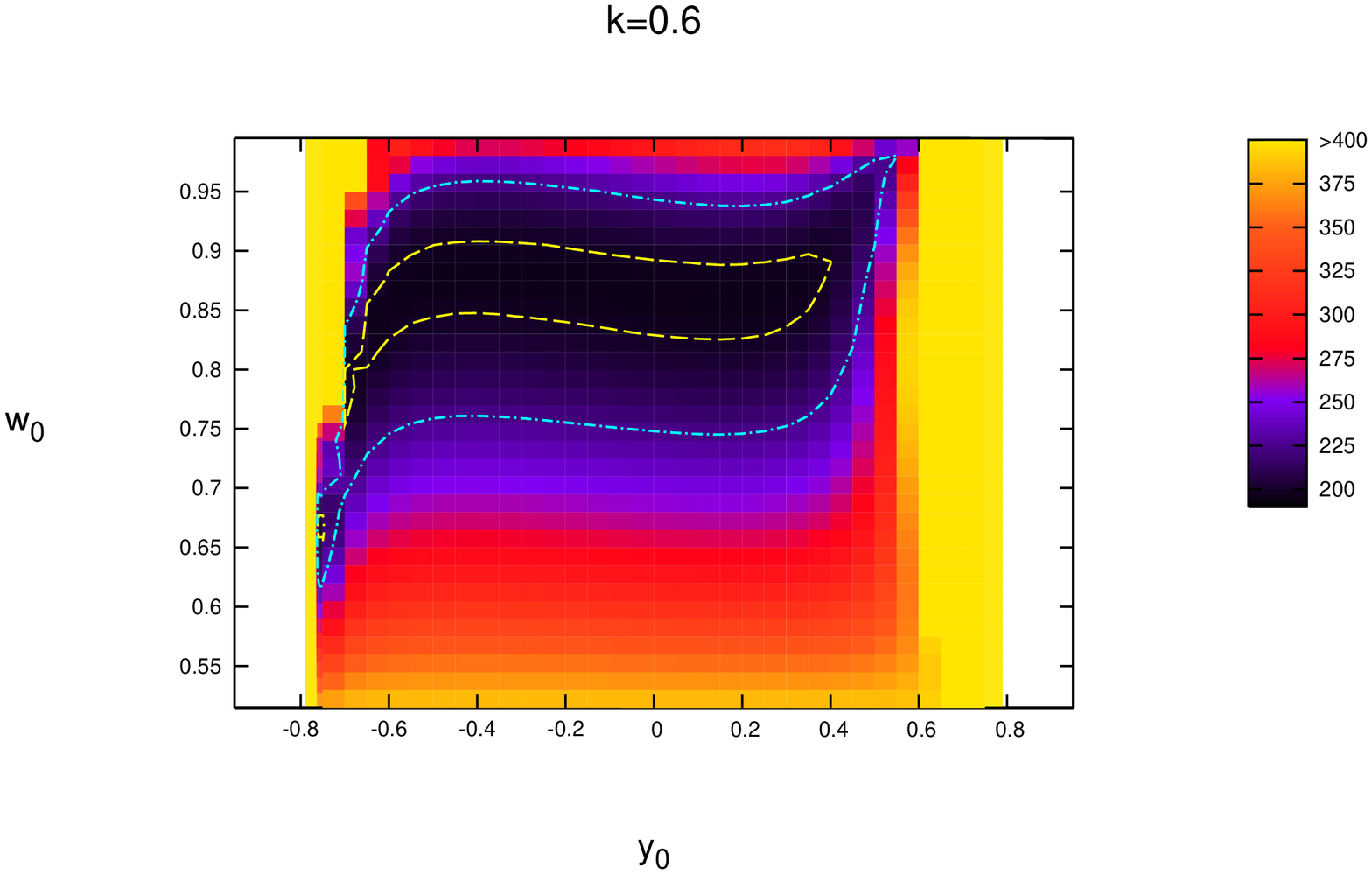}
\caption{(Color online) 
The fit of the luminosity distance vs. redshift for $%
k=-0.4$ (upper left), $-0.2$ (upper right), $0$ (middle left), $0.2$ (middle
right), $0.4$ (lower left), $0.6$ (lower right),
in the parameter plane ($y_{0}$, $w_{0}=1/\left( 1+s_{0}^{2}\right) $).
The white areas represent
regions where the bounds on the model are not satisfied. The contours refer
to the $68.3\%$ (1$\sigma$) and $95.4\%$ (2$\sigma$) confidence levels. For increasing values of $%
\left\vert k\right\vert <1$\thinspace\ the well-fitting regions are
increasingly smaller. The colour code for $\chi^2$ is indicated on the vertical stripes.
}
\label{Fig2}
\end{figure}

\end{widetext}

The dynamical system (\ref{system1})-(\ref{system2}) has three fixed points:
the node (\ref{crit}) and the two saddle points with coordinates 
\begin{equation}
T_1 = \frac{2}{3\sqrt{(1+k)\Lambda}} \mathrm{arccos}\sqrt{\frac{1-k}{1+k}},\
s_1 = 0,  \label{T1}
\end{equation}
and, respectively, 
\begin{equation}
T_2 = \frac{2}{3\sqrt{(1+k)\Lambda}}\left(\pi - \mathrm{arccos}\sqrt{\frac{%
1-k}{1+k}}\right),\ s_2 = 0,  \label{T2}
\end{equation}
which give rise to an unstable de Sitter regime with Hubble parameter $H_1 = 
\sqrt{\frac{(1+k)\Lambda}{2\sqrt{k}}} > H_0$.

The most striking feature of the model under consideration with $k>0$
consists in the fact that now the cosmological trajectories do cross the
corners of the rectangle (\ref{rectangle}),(\ref{interval}). Indeed, the
direct analysis of the system of differential equations in the vicinity of
the points $P,Q,Q^{\prime }$ and $P^{\prime }$ (see Fig. \ref{Fig1}) shows
that these points are not singular points of the system \cite{we-tach}.
Moreover, there is no cosmological singularity in these points \cite{we-tach}%
. That means that the cosmological evolutions must be continued through
them. An apparent obstacle to such a continuation is the fact that the
expression under the square root in the formula for the potential (\ref%
{poten}) changes sign when $T$ becomes smaller than $T_{3}$ or greater than $%
T_{4}$. However, the expression under the square root for the kinetic term $%
\sqrt{1-s^{2}}$ also changes sign at the same time. Then, since the
Lagrangian of the theory is the product of these square roots, these
simultaneous changes of sign leave the Lagrangian and the corresponding
expressions for the energy density (\ref{energy}) and the pressure (\ref%
{pressure}) real. The equation of motion for the tachyon field (\ref%
{eq-of-motion}) also conserves its form. The sign, which we prescribe for
the product (or for the ratio) of the square roots is uniquely determined by
the Friedmann equation. In analyzing the behavior of our dynamical system in
the regions where $|s|>1$ it is convenient to use the new potential 
\begin{eqnarray}
&&W(T)=\frac{\Lambda }{\sin ^{2}\left( \frac{3}{2}\sqrt{\Lambda (1+k)}%
T\right) }  \notag \\
&&\times \sqrt{(1+k)\cos ^{2}\left( \frac{3}{2}\sqrt{\Lambda (1+k)}T\right)
-1},  \label{poten-new}
\end{eqnarray}%
and to substitute in all expressions the term $1-s^{2}$ by $s^{2}-1$. In
doing so the energy density and pressure have the form 
\begin{equation}
\varepsilon =\frac{W(T)}{\sqrt{s^{2}-1}}  \label{en-new}
\end{equation}%
and 
\begin{equation}
p=W(T)\sqrt{s^{2}-1},  \label{pressure-new}
\end{equation}%
being both positive.

The procedure of continuation of the trajectories through the corners of the
rectangle is described in detail in \cite{we-tach}. Here, for the
convenience of the reader we reproduce the phase portrait of the dynamical
system from \cite{we-tach} with some brief comments. The rectangle in the
phase space $(T,s)$ should be complemented by four infinite stripes (see
Fig. \ref{Fig1}). The left upper stripe (the right lower stripe) corresponds
to the initial stages of the cosmological evolution, while the right upper
stripe (the left lower stripe) corresponds to the final stages. There are
five classes of qualitatively different cosmological trajectories. The
trajectories belonging to classes I and II end their evolution with an
infinite de Sitter expansion, while the trajectories of classes III, IV and
V encounter a Big Brake singularity. The curves $\sigma ,\xi ,\tau ,\psi $
and $\chi $ are separatrices, dividing different classes of trajectories.

We end this section with the following remark. Like the other tachyon or DBI
cosmological models (for example, models displaying the power-law or
exponential potentials) the model based on potential (\ref{poten}) possesses
a wide class of cosmological evolutions ending up in an infinite accelerated
expansion. In addition, for small values of $T$, this potential behaves as $%
1/T^{2}$, a behavior which has been widely studied in the literature. So
far, so good. On the other hand, because of the more complicated structure
of the potential (\ref{poten}), our model exhibits another class of
trajectories with a qualitatively very different behavior and, in our
opinion, this is precisely the feature which makes it particularly
interesting.

\section{The tachyon cosmological model and comparison with supernovae type
Ia observational data}

In this section we select, at the confidence level of 1$\sigma $, and for a
given choice of values of the parameter $k$ the set of initial conditions ($%
z=0$) for the system (\ref{system1})--(\ref{system2}), which are compatible
with the supernovae type Ia data taken from paper \cite{SN2007}. To this
purpose, for the numerical analysis of the model it is convenient to rescale
the relevant variables introducing the following dimensionless quantities: 
\begin{equation}
\hat{H}=\frac{H}{H_{0}},\,\hat{V}=\frac{V}{H_{0}^{2}},\,\Omega _{\Lambda }=%
\frac{\Lambda }{H_{0}^{2}},\,\hat{T}=H_{0}T,  \label{new-var}
\end{equation}%
where $H_{0}$ is the present value of the Hubble parameter $H_{0}=H(z=0)$.
In addition we find it convenient to replace the variable $T$ with the new
variable 
\begin{equation*}
y=\cos \left( \frac{3}{2}\sqrt{\Omega _{\Lambda }(1+k)}\hat{T}\right) .
\end{equation*}%
also to switch from the time derivative to the derivative with respect to
the redshift $z$: 
\begin{equation}
\frac{d}{dt}=-H(1+z)\frac{d}{dz},  \label{der-z}
\end{equation}%
and denote $d/dz$ with a prime.

Then, the system of equations (\ref{system1})--(\ref{system2}) in terms of
the new variables $\hat{H},~s,~y$ (all depending on $z$) becomes: 
\begin{eqnarray}
\hat{H}^{2} &=&\frac{\hat{V}}{\left( 1-s^{2}\right) ^{1/2}}\ ,
\label{Hubble2} \\
s &=&\frac{2y^{\prime }\left( 1+z\right) \hat{H}}{3\sqrt{\ \Omega _{\Lambda
}\left( 1+k\right) (1-y^{2})}}\ , \\
\left( 1+z\right) \hat{H}s^{\prime } &=&3\sqrt{\widehat{V}}\left(
1-s^{2}\right) ^{3/4}s  \notag \\
&&+\left( 1-s^{2}\right) \frac{\hat{V}_{,\hat{T}}}{\hat{V}},  \label{system}
\end{eqnarray}%
where $\hat{V}$ and $\hat{V}_{,T}$ are given by 
\begin{eqnarray}
\hat{V} &=&\frac{\ \Omega _{\Lambda }\left[ 1-\left( 1+k\right) y^{2}\right]
^{1/2}}{1-y^{2}}\ ,  \label{V} \\
\hat{V}_{,\hat{T}} &=&\frac{3\Omega _{\Lambda }\sqrt{\ \Omega _{\Lambda
}\left( 1+k\right) }y\left[ k-1+\left( 1+k\right) y^{2}\right] }{%
2(1-y^{2})^{3/2}\left[ 1-\left( 1+k\right) y^{2}\right] ^{1/2}}\ .
\label{VT}
\end{eqnarray}

Since $\hat{H}^{2}\left( 0\right) =1$, the present day values of the
variables $s$ and $y$ satisfy the constraint 
\begin{equation*}
s\left( 0\right) =\pm \sqrt{1-\frac{\ \Omega _{\Lambda }^{2}\left[ 1-\left(
1+k\right) y\left( 0\right) ^{2}\right] }{\left[ 1-y^{2}\left( 0\right) %
\right] ^{2}}}\ .
\end{equation*}%
We can avoid double coverage of the parameter space (the model being
invariant under the simultaneous change of signs $y_{0}\rightarrow -y_{0}$
and $s_{0}\rightarrow -s_{0}$) by replacing $s_{0}$ by the new variable 
\begin{equation}
w_{0}=\frac{1}{1+s_{0}^{2}}~.  \label{w0}
\end{equation}

The luminosity distance function for a flat Friedmann universe 
\begin{equation}
d_{L}\left( z\right) =\left( 1+z\right) \int_{0}^{z}\frac{dz^{\ast }}{%
H\left( z^{\ast }\right) }\ 
\end{equation}%
gives for the dimensionless luminosity distance $\hat{d}_{L}=H_{0}d_{L}$ the
equation 
\begin{equation}
\left( \frac{\hat{d}_{L}}{1+z}\right) ^{\prime }=\frac{1}{\hat{H}}\ .
\label{dLz}
\end{equation}

We are now in a position to compare our model with the Supernovae type Ia
data \cite{SN2007}.

Following Ref. \cite{DicusRepko} we introduce the distance modulus type
quantity $5\log _{10}\hat{d}_{L}\left( z\right) +M$, with $M$ a constant
offset between the data and the theoretical expression. The comparison
involves computing 
\begin{equation}
\chi ^{2}=\sum_{i=1}^{N}\frac{1}{\sigma _{i}^{2}}\left[ 5\log _{10}\hat{d}%
_{L}^{\exp }\left( z_{i}\right) -M-5\log _{10}\hat{d}_{L}\left( z_{i}\right) %
\right] ^{2}\ ,
\end{equation}%
where the sum is over the supernovae in the data set and $\sigma _{i}$ are
the experimental errors in $5\log _{10}\hat{d}_{L}^{\exp }\left(
z_{i}\right) $. The distance luminosity function $\hat{d}_{L}\left( z\right) 
$ depends on the initial condition $y_{0}=y\left( 0\right) $ and $%
s_{0}=s\left( 0\right) $. We minimize this expression with respect to $M$
obtaining 
\begin{equation}
M=\frac{L}{D}\ ,
\end{equation}%
with%
\begin{eqnarray}
L &=&\sum_{i=1}^{N}\frac{1}{\sigma _{i}^{2}}\left[ 5\log _{10}\hat{d}%
_{L}^{\exp }\left( z_{i}\right) -5\log _{10}\hat{d}_{L}\left( z_{i}\right) %
\right] \ , \\
D &=&\sum_{i=1}^{N}\frac{1}{\sigma _{i}^{2}}\ .
\end{eqnarray}

In Table \ref{ys} are listed the values $y_{j},~j=1,2,3,4$, of the variable $%
y$ corresponding to the values $T_{j}$ of the variable $T$ given in formulas
(\ref{T3})--(\ref{T2}) for the chosen positive values of $k$. 
\begin{table}[th]
\caption{The values of $y_{j}$ (corresponding to the $T_{j}$) for some
positive values of $k$.}
\label{ys}
\begin{center}
\begin{tabular}{l|lll}
$k$ & $0.2$ & $0.4$ & $0.6$ \\ \hline
$y_{1,2}$ & $\pm 0.816$ & $\pm 0.655$ & $\pm 0.500$ \\ 
$y_{3,4}$ & $\pm 0.913$ & $\pm 0.845$ & $\pm 0.791$%
\end{tabular}%
\end{center}
\end{table}

Since the expansion of the present day universe is accelerated the pressure
is negative, hence $|s_{0}|<1$. Therefore, the initial point in the phase
diagram ($T,s$) should lie inside the rectangle ($T_{3}<T<T_{4},|s|<1$),
(see Fig. \ref{Fig1}). Thus the bounds on the model are not satisfied in the
ranges $y_{0}<y_{4}$ and $y_{0}>y_{3}$.

In Fig \ref{Fig2} we represent the values of $\chi ^{2}$ in the parameter
plane of the initial conditions $\left( y_{0}=y(0),w_{0}=w(0)\right) $, for
the choices $k=0$, $\pm 0.2$,$~\pm 0.4$ and $0.6$. The contours represent
the $68.3$ (1$\sigma $) and $95.4$ (2$\sigma $) confidence levels and the
white areas are unallowed regions.

\section{Future cosmological evolutions}

In this section, in order to investigate the possible futures of the
universe within the tachyon cosmological model, we evolve numerically the
model forward in time starting from the parameter range ($w_{0},y_{0}$) of
initial conditions for which the fitting with the supernovae data is within 1%
$\sigma $ ($68.3\%$) confidence level. We do this by numerical integration
of equations of motion from $z=0$ towards negative values of $z$.

\begin{widetext}

\begin{figure}[ht]
\vskip 0.8cm %
\includegraphics[height=7cm, angle=270]{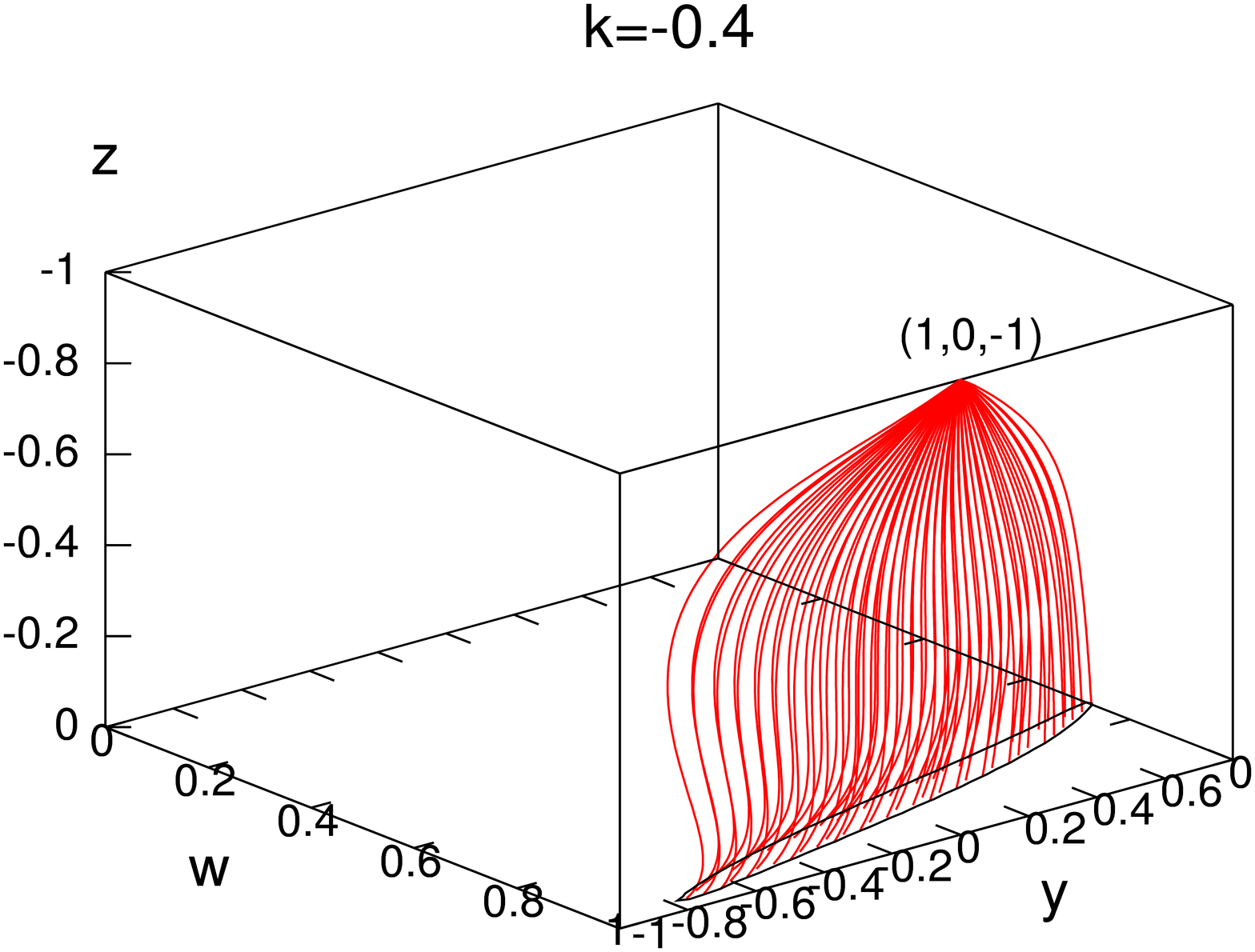} %
\hskip 2cm %
\includegraphics[height=7cm,angle=270]{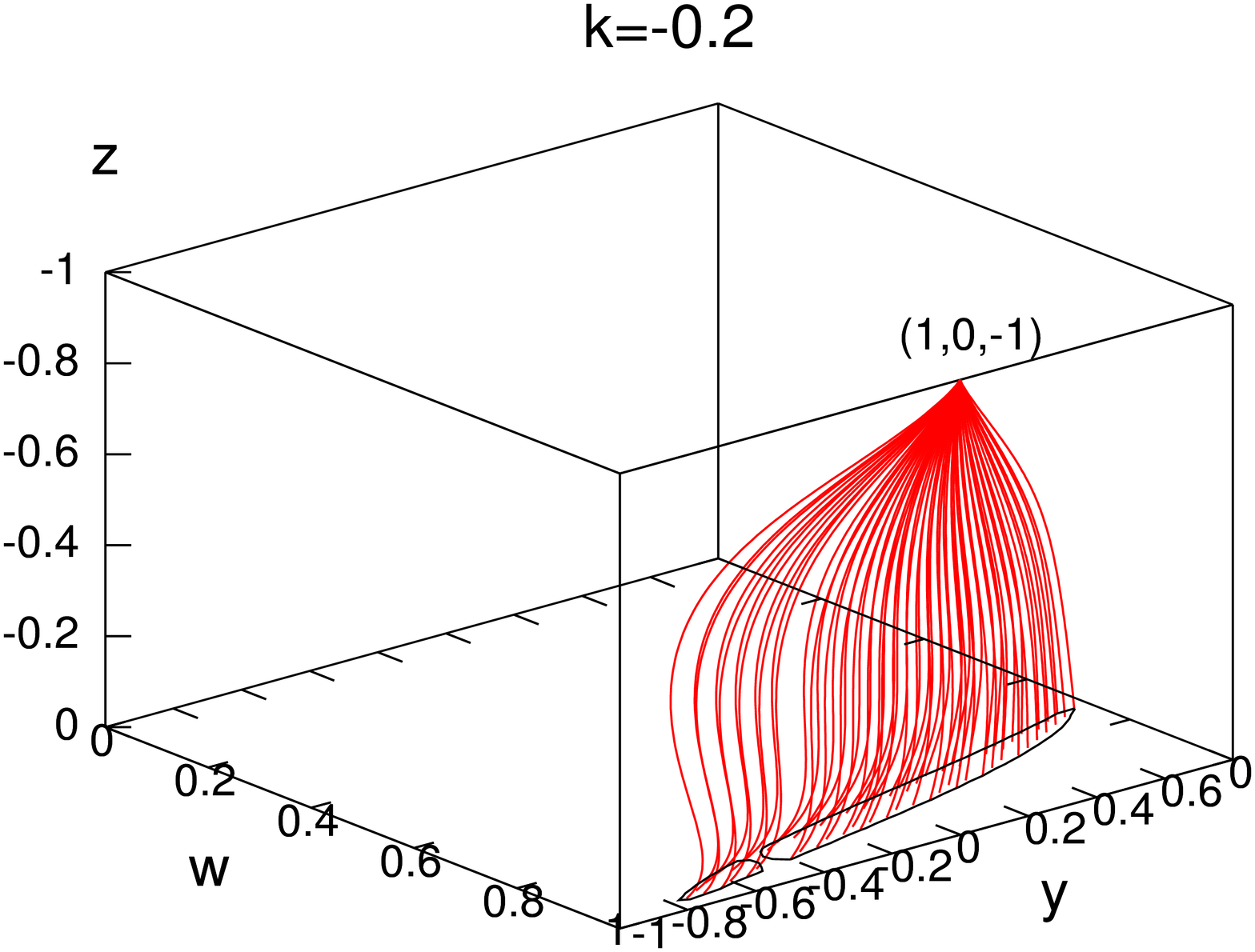}
\newline
\vskip 0.8cm
\includegraphics[height=7cm,
angle=270]{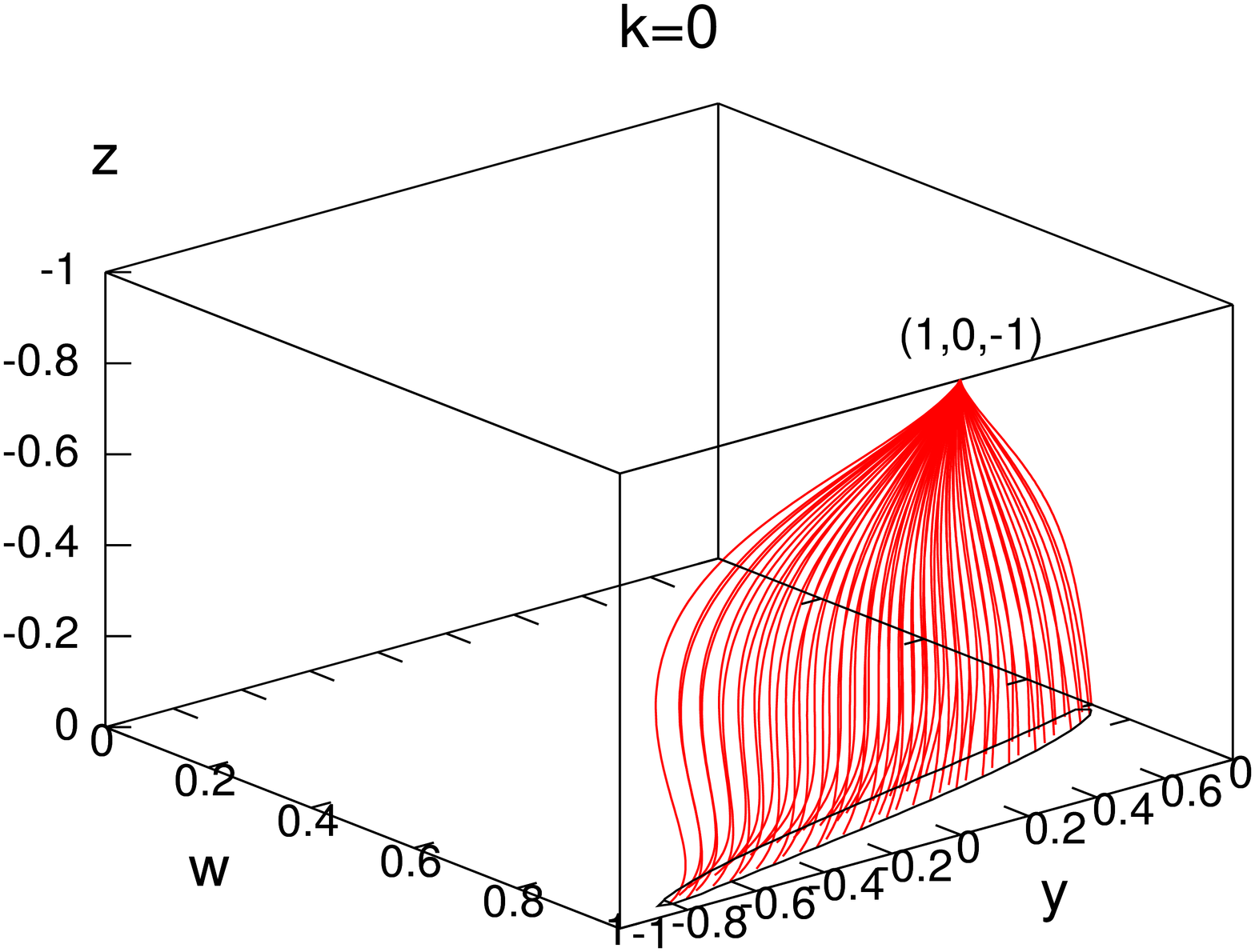} \hskip 2cm %
\includegraphics[height=7cm,angle=270]{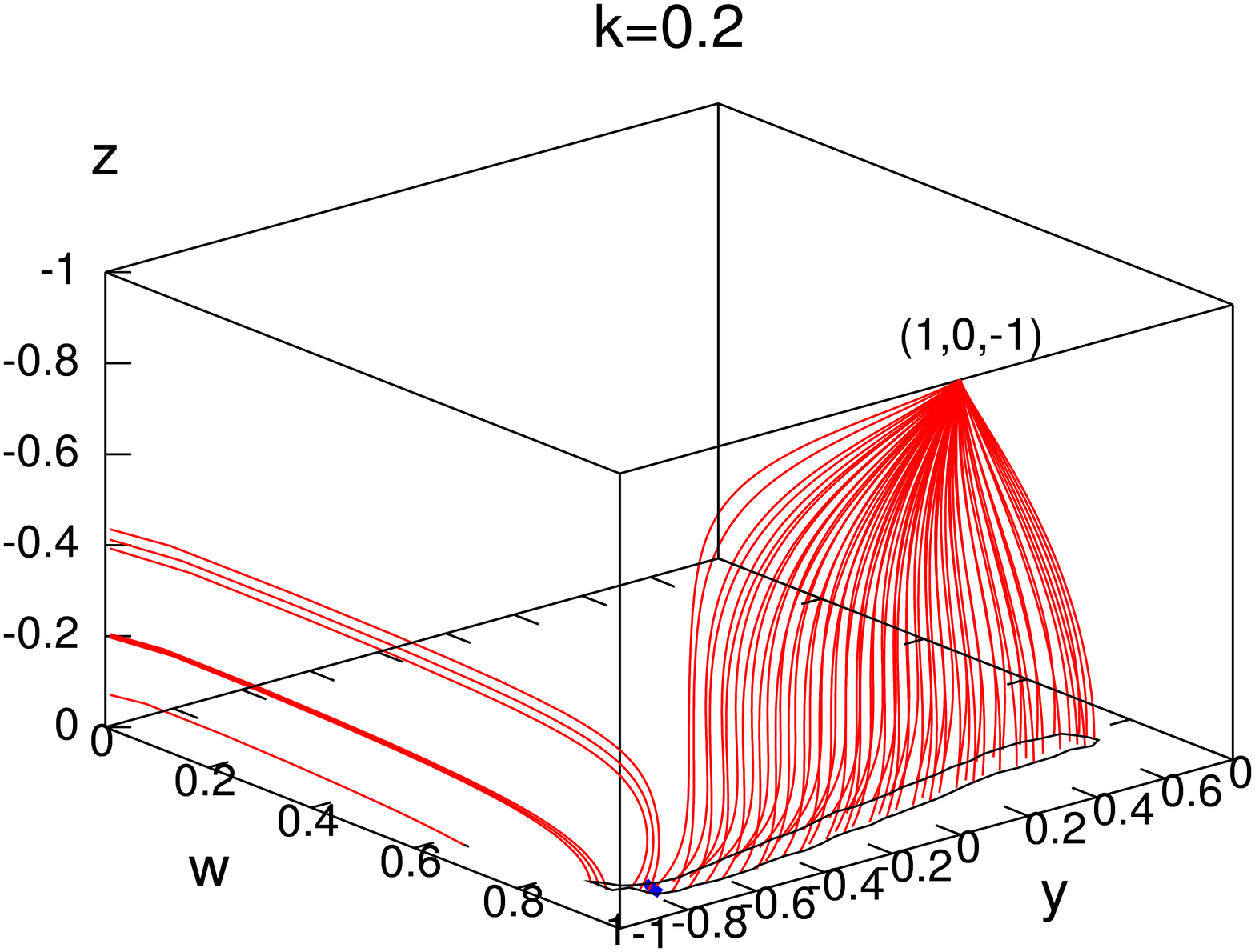}
\newline
\vskip 0.8cm
\includegraphics[height=7cm,
angle=270]{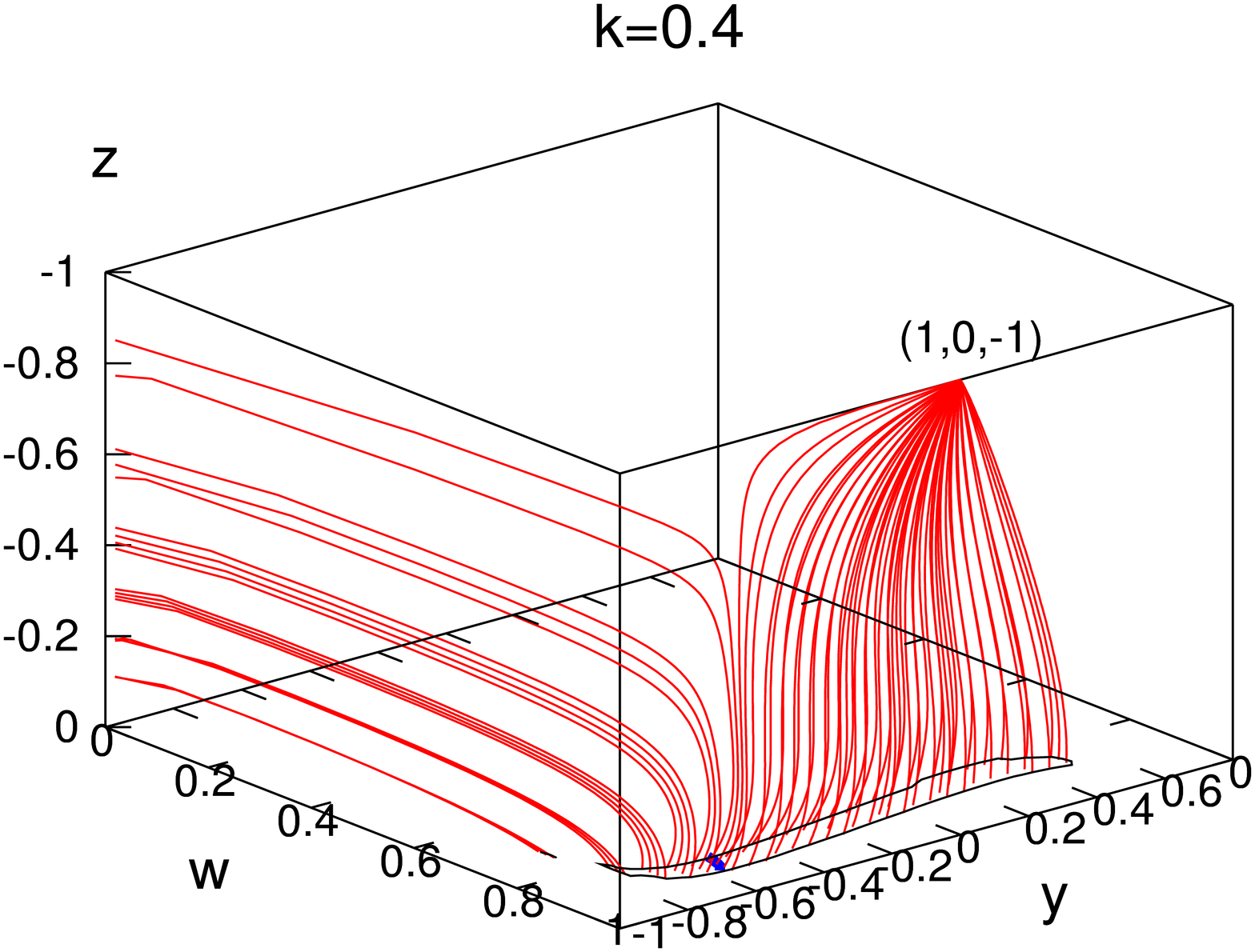} \hskip 2cm %
\includegraphics[height=7cm,angle=270]{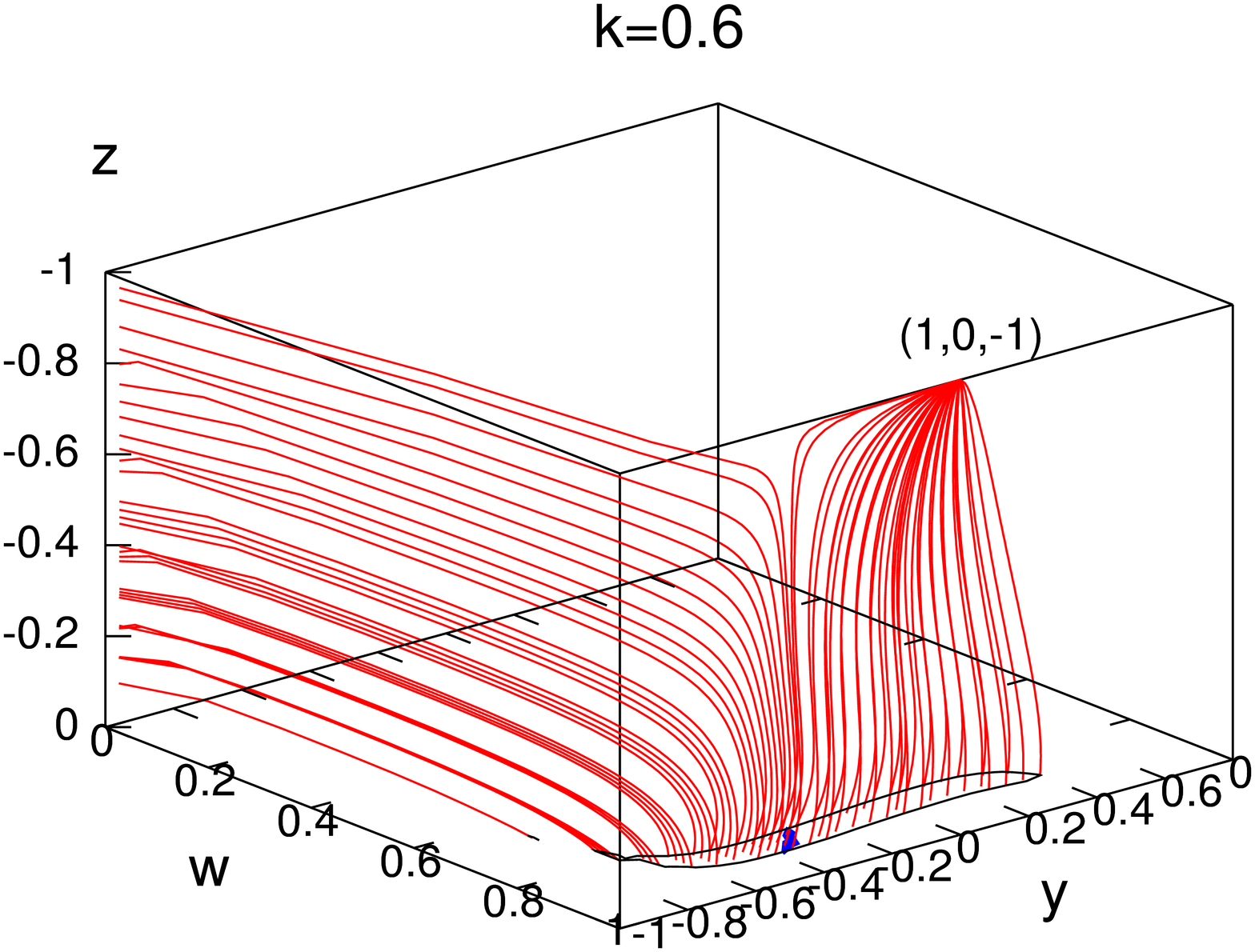}
\newline
\vskip 0.6cm
\caption{(Color online) The future evolution of those universes, which are
in a $68.3\%$ confidence level fit with the supernova data. The 1$\protect%
\sigma $ contours (black lines in the $z=0$ plane) are from Fig \protect\ref%
{Fig2} (the parameter plane $\left( y_{0},w_{0}\right) $ is the $z=0$ plane
here). The sequence of figures and the values of $k$ are the same as on Fig. \protect\ref{Fig2}.
The short and thick (blue) line in the plane of initial conditions separates the 1$%
\protect\sigma $ parameter ranges for which the universe evolves into a de
Sitter regime or towards the Big Brake singularity. Future evolutions
towards the Big Brake singularity of the universes selected by the
comparison with supernovae data become more frequent with increasing $k$. }
\label{Fig3}
\end{figure}

\end{widetext}

The results of these computations, corresponding to the six values of $k$
chosen earlier, are displayed in Fig. \ref{Fig3} in the space ($%
w=(1+s^{2})^{-1},y,z$). The evolution curves start from the allowed region ($%
w_{0},y_{0}$) in the plane $z=0$. The final de Sitter state is characterized
by the point ($w_{dS}=1,y_{dS}=0,z_{dS}=-1$), the Big Brake final state by
points ($w_{BB}=0,-1<y_{BB}<0,-1<z_{BB}<0$).

Whereas all trajectories with $k\leq 0$ end up eventually into the de Sitter
state, those with $k>0$ can either evolve into the de Sitter state or into
the Big Brake state, depending on the particular initial condition ($%
w_{0},y_{0}$). The fraction of curves eventually meeting a Big Brake
increases with increasing $k$. This is clearly seen in Fig. \ref{Fig3} from
the relative sizes of the 1$\sigma $ subdomains belonging to these two
regimes, which are separated by a line.

For all future evolutions encountering a Big Brake singularity we have
computed the actual time $t_{BB}$ it will take to reach the singularity,
measured from the present moment $z = 0$, using the equation $\left(
H_{0}t\right) ^{\prime }=-\hat{H}^{-1}\left( 1+z \right)^{-1}$). The results
are shown in Tables \ref{Table1}-\ref{Table3}. In the tables the parameter
values at which the pressure turns from negative to positive are also
displayed.

\begin{table}[t]
\caption{Properties of the tachyonic universes with $k=0.2$ which (a) are
within 1$\protect\sigma $ confidence level fit with the type Ia supernova
data and (b) evolve into a Big Brake singularity. Columns (1) and (2)
represent a grid of values of the allowed model parameters. Columns (3) and
(4): the redshift $z_{\ast }$ and time $t_{\ast }$ at the future tachyonic
crossing (when $s=1$ and the pressure becomes positive). Columns (5) and
(6): the redshift $z_{BB}$ and time $t_{BB}$ necessary to reach the Big
Brake. The former indicates the relative size of the universe when it
encounters the Big Brake. (The values of $t_{\ast }$ and $t_{BB}$ were
computed with the Hubble parameter $H_{0}=73$ km/s/Mpc.)}
\label{Table1}
\begin{center}
\begin{tabular}{c|c|c|c|c|c}
$y_{0}$ & $w_{0}$ & $z_{\ast }$ & $t_{\ast }\left( 10^{9}yrs\right) $ & $%
z_{BB}$ & $t_{BB}\left( 10^{9}yrs\right) $ \\ \hline
$-0.90$ & $0.635$ & $-0.024$ & $0.3$ & $-0.068$ & $1.0$ \\ 
$-0.85$ & $0.845$ & $-0.158$ & $2.4$ & $-0.194$ & $3.1$ \\ 
$-0.85$ & $0.860$ & $-0.162$ & $2.4$ & $-0.198$ & $3.1$ \\ 
$-0.85$ & $0.875$ & $-0.166$ & $2.5$ & $-0.201$ & $3.2$ \\ 
$-0.80$ & $0.890$ & $-0.363$ & $6.2$ & $-0.390$ & $6.9$ \\ 
$-0.80$ & $0.905$ & $-0.384$ & $6.7$ & $-0.409$ & $7.3$ \\ 
$-0.80$ & $0.920$ & $-0.408$ & $7.2$ & $-0.432$ & $7.9$%
\end{tabular}%
\end{center}
\end{table}

\begin{table}[t]
\caption{As in Table \protect\ref{Table1}, for $k=0.4$.}
\label{Table2}
\begin{center}
$%
\begin{tabular}{c|c|c|c|c|c}
$y_{0}$ & $w_{0}$ & $z_{\ast }$ & $t_{\ast }\left( 10^{9}yrs\right) $ & $%
z_{BB}$ & $t_{BB}\left( 10^{9}yrs\right) $ \\ \hline
$-0.80$ & $0.710$ & $-0.059$ & $0.8$ & $-0.106$ & $1.6$ \\ 
$-0.80$ & $0.725$ & $-0.059$ & $0.8$ & $-0.105$ & $1.6$ \\ 
$-0.80$ & $0.740$ & $-0.060$ & $0.8$ & $-0.105$ & $1.6$ \\ 
$-0.75$ & $0.815$ & $-0.144$ & $2.1$ & $-0.184$ & $2.9$ \\ 
$-0.75$ & $0.830$ & $-0.147$ & $2.2$ & $-0.187$ & $3.0$ \\ 
$-0.75$ & $0.845$ & $-0.150$ & $2.2$ & $-0.189$ & $3.0$ \\ 
$-0.70$ & $0.845$ & $-0.241$ & $3.8$ & $-0.276$ & $4.6$ \\ 
$-0.70$ & $0.860$ & $-0.248$ & $4.0$ & $-0.282$ & $4.7$ \\ 
$-0.70$ & $0.875$ & $-0.256$ & $4.1$ & $-0.290$ & $4.9$ \\ 
$-0.70$ & $0.890$ & $-0.264$ & $4.2$ & $-0.298$ & $5.0$ \\ 
$-0.65$ & $0.860$ & $-0.358$ & $6.2$ & $-0.387$ & $7.0$ \\ 
$-0.65$ & $0.875$ & $-0.372$ & $6.5$ & $-0.400$ & $7.2$ \\ 
$-0.65$ & $0.890$ & $-0.388$ & $6.8$ & $-0.415$ & $7.6$ \\ 
$-0.65$ & $0.905$ & $-0.406$ & $7.2$ & $-0.432$ & $8.0$ \\ 
$-0.60$ & $0.875$ & $-0.521$ & $10$ & $-0.542$ & $11$ \\ 
$-0.60$ & $0.890$ & $-0.551$ & $11$ & $-0.571$ & $12$ \\ 
$-0.60$ & $0.905$ & $-0.587$ & $12$ & $-0.605$ & $13$ \\ 
$-0.55$ & $0.875$ & $-0.756$ & $19$ & $-0.766$ & $20$ \\ 
$-0.55$ & $0.890$ & $-0.837$ & $25$ & $-0.845$ & $26$%
\end{tabular}%
$%
\end{center}
\end{table}

\begin{table}[t]
\caption{As in Table \protect\ref{Table1}, for $k=0.6$. The evolutions into
a Big Brake Singularity compatible with supernova observations are more
numerous with increasing $k$.}
\label{Table3}
\begin{center}
$%
\begin{tabular}{c|c|c|c|c|c}
$y_{0}$ & $w_{0}$ & $z_{\ast }$ & $t_{\ast }\left( 10^{9}yrs\right) $ & $%
z_{BB}$ & $t_{BB}\left( 10^{9}yrs\right) $ \\ \hline
$-0.75$ & $0.665$ & $-0.039$ & $0.5$ & $-0.088$ & $1.4$ \\ 
$-0.70$ & $0.755$ & $-0.098$ & $1.4$ & $-0.145$ & $2.3$ \\ 
$-0.70$ & $0.770$ & $-0.100$ & $1.5$ & $-0.145$ & $2.3$ \\ 
$-0.70$ & $0.785$ & $-0.101$ & $1.5$ & $-0.146$ & $2.3$ \\ 
$-0.70$ & $0.800$ & $-0.102$ & $1.5$ & $-0.146$ & $2.3$ \\ 
$-0.65$ & $0.815$ & $-0.168$ & $2.6$ & $-0.209$ & $3.4$ \\ 
$-0.65$ & $0.830$ & $-0.171$ & $2.6$ & $-0.212$ & $3.4$ \\ 
$-0.65$ & $0.845$ & $-0.175$ & $2.7$ & $-0.215$ & $3.5$ \\ 
$-0.60$ & $0.830$ & $-0.240$ & $3.9$ & $-0.277$ & $4.7$ \\ 
$-0.60$ & $0.845$ & $-0.247$ & $4.0$ & $-0.283$ & $4.8$ \\ 
$-0.60$ & $0.860$ & $-0.254$ & $4.1$ & $-0.289$ & $4.9$ \\ 
$-0.60$ & $0.875$ & $-0.261$ & $4.2$ & $-0.296$ & $4.0$ \\ 
$-0.55$ & $0.845$ & $-0.325$ & $5.5$ & $-0.357$ & $6.3$ \\ 
$-0.55$ & $0.860$ & $-0.335$ & $5.7$ & $-0.366$ & $6.5$ \\ 
$-0.55$ & $0.875$ & $-0.347$ & $5.9$ & $-0.377$ & $6.7$ \\ 
$-0.55$ & $0.890$ & $-0.359$ & $6.2$ & $-0.389$ & $7.0$ \\ 
$-0.50$ & $0.845$ & $-0.411$ & $7.5$ & $-0.439$ & $8.3$ \\ 
$-0.50$ & $0.860$ & $-0.427$ & $7.8$ & $-0.453$ & $8.6$ \\ 
$-0.50$ & $0.875$ & $-0.444$ & $8.2$ & $-0.469$ & $9.0$ \\ 
$-0.50$ & $0.890$ & $-0.463$ & $8.6$ & $-0.488$ & $9.4$ \\ 
$-0.45$ & $0.860$ & $-0.533$ & $10$ & $-0.554$ & $11$ \\ 
$-0.45$ & $0.875$ & $-0.557$ & $11$ & $-0.577$ & $12$ \\ 
$-0.45$ & $0.890$ & $-0.584$ & $12$ & $-0.603$ & $13$ \\ 
$-0.45$ & $0.905$ & $-0.616$ & $13$ & $-0.633$ & $14$ \\ 
$-0.40$ & $0.860$ & $-0.658$ & $15$ & $-0.673$ & $16$ \\ 
$-0.40$ & $0.875$ & $-0.693$ & $16$ & $-0.707$ & $17$ \\ 
$-0.40$ & $0.890$ & $-0.733$ & $18$ & $-0.745$ & $19$ \\ 
$-0.40$ & $0.905$ & $-0.779$ & $21$ & $-0.789$ & $22$ \\ 
$-0.35$ & $0.860$ & $-0.814$ & $23$ & $-0.822$ & $24$ \\ 
$-0.35$ & $0.875$ & $-0.865$ & $28$ & $-0.872$ & $29$ \\ 
$-0.35$ & $0.890$ & $-0.927$ & $36$ & $-0.930$ & $37$ \\ 
$-0.30$ & $0.845$ & $-0.955$ & $43$ & $-0.957$ & $44$%
\end{tabular}%
$%
\end{center}
\end{table}

Finally we have evolved numerically backward in time some of the
trajectories crossing the 1$\sigma $ domain, until they reached one of the
Big Bang singularities of the model. All trajectories we have checked
originate from the singularity at $|s|=1$. In other words, they start from
the horizontal boundaries of the rectangle in the phase plane ($T,s$), and
depending on whether they evolve into an infinite de Sitter expansion or
reach the Big Brake singularity, they belong to either type II or III.

\section{Concluding Remarks}

In this paper we have shown that the tachyon cosmological model of Ref. \cite%
{we-tach} allows for a consistent set of trajectories which are compatible
with the supernovae type Ia data.

We have found that, among these, for positive values of the parameter $k$ of
the model, there is a subset of evolutions which end up into a Big Brake
singularity and, for the latter, we have computed the relevant Big Brake
parameters $z_{BB}$ and $t_{BB}$.

The compatibility of cosmological evolutions possessing soft cosmological
singularities with the supernovae type Ia data was studied in \cite{soft1}.
Curiously, it was found in Ref. \cite{soft1} that a sudden singularity may
take place in already a very close future, even less then 10 million years.
However this analysis was purely kinematical, and we also note that the
parameters in our model (as given by the tachyonic dynamics) near the Big
Brake singularity fall outside the range considered in \cite{soft1}. The
problem of stability of a cosmological evolution in the vicinity of such
singularities was studied in \cite{soft2}.

Finally, we may ask why the model proposed in \cite{we-tach} is worth
studying. First, the soft (sudden) cosmological singularity of the Big Brake
type arises in our model in a very natural way as a particular class of
solutions of the dynamical system (\ref{system1})--(\ref{system2}). Second,
the model has another interesting feature. A subtle interplay between
geometry and matter, induces a change of the very nature of the latter: it
transforms from tachyon into a ``pseudo-tachyon'' field (see \cite{we-tach}
for details). We point out that a similar effect was observed also in
scalar-phantom cosmological models \cite{we-phan}. Phenomena of this kind
represent a distinguishing feature of general relativity \cite{gen}: the
requirement of self-consistency of Einstein equations can impose the form of
the equations of motion for the matter.

Thus, in spite of being a toy model, the tachyon cosmological model \cite%
{we-tach} can serve as a prototype of realistic (i.e. compatible with
observational data) cosmological models which may lead to a final fate of
the Universe, different from the infinite quasi - de Sitter expansion of the 
$\Lambda $CDM model. What will actually happen in the future is left to our
far away descendants to experience!

\section*{Acknowledgements}

We thank Gy. Szab\'{o} for discussions in the early stages of this project.
We are grateful to J.D. Barrow and M.P. D\c{a}browski for useful
correspondence. Z.K. was supported by the OTKA grant 69036; L.\'{A}.G. was
supported by the OTKA grant 69036, the London South Bank University Research
Opportunities Fund and the Pol\'{a}nyi Program of the Hungarian National
Office for Research and Technology (NKTH); A.K. was partially supported by
RFBR grant No. 08-02-00923 and by the grant LSS-4899.2008.2.

\end{document}